\documentclass{article}
\usepackage[nonatbib,final]{nips_2017}

\usepackage[utf8]{inputenc} % allow utf-8 input
\usepackage[T1]{fontenc}    % use 8-bit T1 fonts
\usepackage{hyperref}       % hyperlinks
\usepackage{url}            % simple URL typesetting
\usepackage{booktabs}       % professional-quality tables
\usepackage{amsfonts}       % blackboard math symbols
\usepackage{nicefrac}       % compact symbols for 1/2, etc.
\usepackage{microtype}      % microtypography
\usepackage[urldate=long,backend=biber,sorting=none,sortcites=true,natbib=true,style=numeric-comp]{biblatex}
\usepackage{graphicx}
\usepackage{import}
\usepackage{float}
\usepackage{subfig}
\usepackage{siunitx}
\usepackage{amsmath}
\usepackage{enumitem}

\DeclareMathOperator{\sign}{sign}
\newcommand{\etal}{\textit{et al}. }

\newcommand{\estimates}{\overset{\scriptscriptstyle\wedge}{=}}

\title{A Connectome Based Hexagonal Lattice Convolutional Network Model of the Drosophila Visual System}

\bibliography{bib.bib}

\graphicspath{{./data/}}

\author{
  Fabian David ~Tschopp\\
  Institute of Neuroinformatics\\
  University of Zurich and ETH Zurich\\
  \texttt{tschopfa@student.ethz.ch} \\
  \And
  Michael B. ~Reiser \\
  Janelia Research Campus \\
  Howard Hughes Medical Instutute \\
  \texttt{reiserm@janelia.hhmi.org} \\
  \AND
  Srinivas C. ~Turaga \\
  Janelia Research Campus \\
  Howard Hughes Medical Instutute \\
  \texttt{turagas@janelia.hhmi.org} \\
}

\makeatletter
\def\blfootnote{\gdef\@thefnmark{}\@footnotetext}
\makeatother

\begin{document}

\maketitle

\begin{abstract}
	What can we learn from a connectome? We constructed a simplified model of the first two stages of the fly visual system, the lamina and medulla. The resulting hexagonal lattice convolutional network was trained using backpropagation through time to perform object tracking in natural scene videos. Networks initialized with weights from connectome reconstructions automatically discovered well-known orientation and direction selectivity properties in T4 neurons and their inputs, while networks initialized at random did not. Our work is the first demonstration, that knowledge of the connectome can enable in silico predictions of the functional properties of individual neurons in a circuit, leading to an understanding of circuit function from structure alone.
\end{abstract}

\blfootnote{Work in progress. \\
Supplemental Material can be found at \href{https://github.com/naibaf7/dvsc\_2017\_supplemental}{https://github.com/naibaf7/dvsc\_2017\_supplemental}}

\section{Introduction}
\label{sec:intro}

Universal function approximation results for artificial neural networks \cite{Csaji2001,Cybenko1992} imply that many possible neural network architectures, with different connectivity matrices and different activation functions, can compute the same function. This suggests that it is difficult to predict the precise neural circuitry underlying a given neural computation. For this reason, we have many different proposals for instance, for how visual motion is computed through elementary motion detectors (EMD) such as Barlow-Levick \cite{Barlow1965} and Hassenstein-Reichardt \cite{Reichardt1961}, and how neural signals might be integrated \cite{Seung1996,Goldman2009}. In contrast, since neural connectivity is generally sparse, knowledge of the connectivity of a neural circuit might constrain its possible computational function \cite{Seung2009,Helmstaedter2013}. Taken together, this suggests that it might be easier to predict function from structure, rather than the other way around.

Recent reconstructions of the first two stages of the drosophila visual system \cite{Takemura2013,Takemura2015,Takemura2017,RiveraAlba2011,Tuthill2013,Tuthill2014} enable us to test this hypothesis that neural circuit function can be predicted from neural circuit structure. We constructed a connectome-based artificial network model of the drosophila visual system and trained this model to track objects in videos of natural scenes. We found that a network model derived from the connectome is able to recapitulate well-known orientation and direction selectivity tuning properties of cell types in the medulla of the circuit. 

% TODO: Remove this before final submission
\newpage

\subsection{Prior work}
Existing work has focused on roughly correlating single cell responses in different brain brain regions to layers in a deep convolutional neural network \cite{Yamins2014,Pinto2009}, but not identify individual cell types and their connectivity. A different hexagonal lattice based model has also been proposed \cite{Givon2016,Lazar2015}, however they do not model any circuitry beyond the photoreceptors, focusing only on the simulation of ommatidia. Specific to the \textit{Drosophila}, mathematical models to fit the physiological motion responses have been suggested \cite{Serbe2016}, however these simpler models were explicitly trained to learn the neural response. Other recent work tries to infer structure from function \cite{Lee2017}, without a connectome, and shows that a deep network can learn the neural responses of retinal ganglion cells when a convolutional network based encoding model is trained on natural images and white noise. They further show that recurrent lateral connections and feedforward inhibition help to reconstruct temporally accurate retinal responses. Physiological experiments \cite{Tuthill2013} suggest that this is also the case for directional motion detection in the \textit{Drosophila}.

\section{Hexagonal lattice convolutional network model}
\label{sec:network_model}
Based on publications of lamina \cite{Tuthill2013,Tuthill2014,RiveraAlba2011} and medulla \cite{Takemura2013,Takemura2015,Takemura2017} connectomes, we modeled a connectome consisting of 43 neuron types. The neurons in these layers have a repeating columnar architecture with one column per ommatidium, each spanning $5^\circ$ of visual angle, and they form a hexagonal lattice \cite{Braitenberg1967,Kirschfeld1967}. Since the published connectomes only correspond to reconstructions of a few columns and not the entire hexagonal lattice, we repeat the locally described connectome in a spatially invariant manner, leading to a hexagonal lattice convolutional network.

\begin{figure}
	\includegraphics[width=0.95\linewidth]{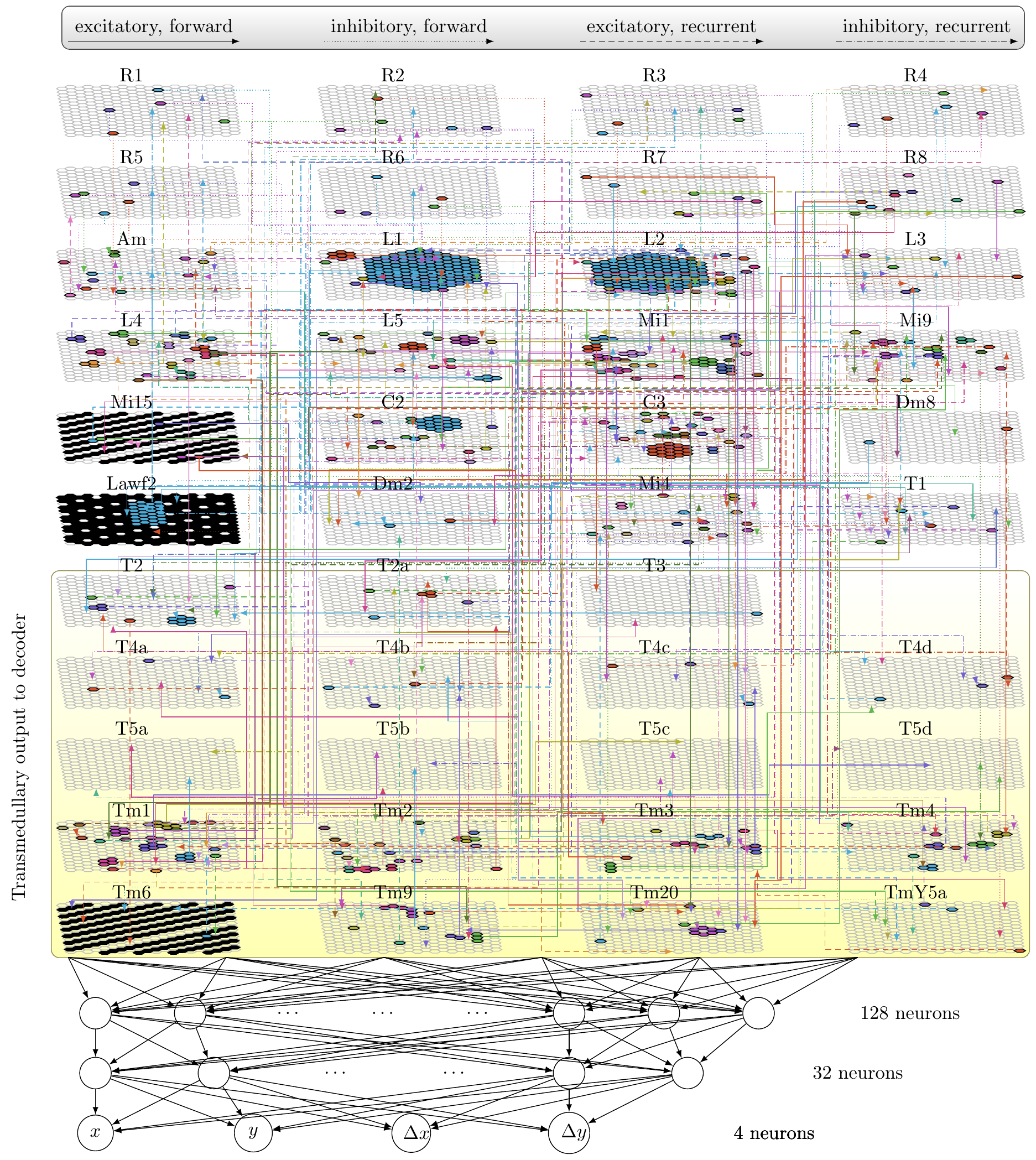}
  	\caption{\textit{Drosophila} visual system connectome-based model with 3 layer decoder.}
	\label{fig:fullnet}
\end{figure}

The network begins with the bundled rhabdomeres R1-R8. We simplified the neural superposition mapping \cite{Kirschfeld1967,Morante2004} by directly associating the inputs with their target column in the lamina and medulla, where they form inhibitory synapses \cite{Hardie1989}. For the lamina, we simulate 10 neuron classes (L1-L5, C2-C3, T1, Lawf2, Am) that are known to contribute to motion processing. L1-L5, C2-C3 and T1 are synperiodic \cite{Takemura2013} and therefore present in each column. Amacrine cells (Am) are multicolumnar. Lawf2 cells are likely presynaptic to most cells in the lamina, have wide receptive fields in the medulla and feedback to the lamina. As there are $\sim 140$ Lawf2 neurons per optic lobe, and each column is innervated by $\sim 5$ Lawf2 cells \cite{Tuthill2014}, we can distribute Lawf2 sparsely in our model (Figure \ref{fig:fullnet}, blacked out neurons are set to be constantly disabled). We had to omit Lawf1, Lai, Lat and glia cells due to a lack of dense reconstructions of their connectivity.

\begin{figure}
	\centering
	\subfloat{
		\includegraphics[scale=0.4]{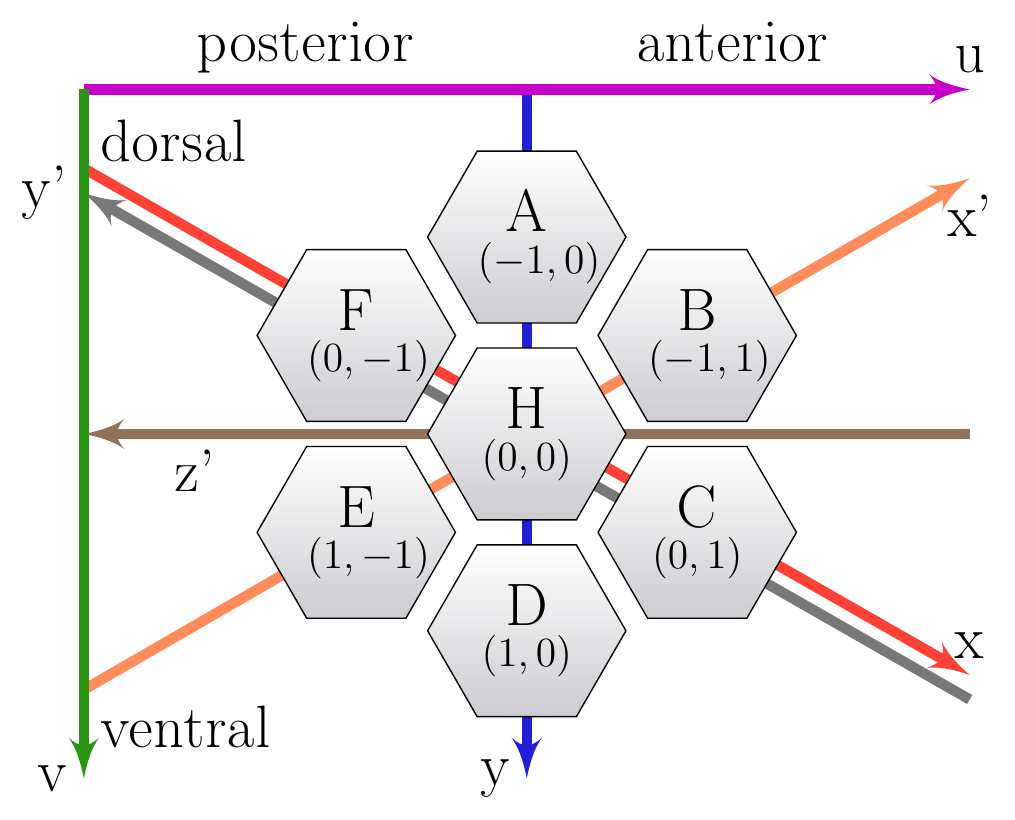}
	}
	\hspace{5mm}
	\subfloat{
		\makebox[85mm][c]{\includegraphics[scale=1.3]{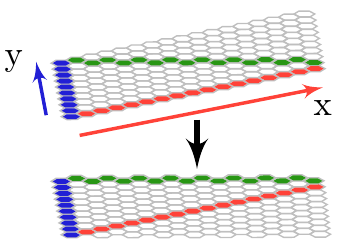}}
	}
	\caption{(left) 7 column structure with 6-neighborhood offsets along the principal $y$ and $x$ axis. The axis $y'$, $x'$ and $z'$ are equivalent to Braitenberg's hexagonal arrangement of ommatidia \cite{Braitenberg1967}. (right) Semantic arrangement of the hexagonal lattice of size $h\times w$ on the GPU as a 2D tensor. Additional border conditions at $y+\frac{1}{2}\cdot x = h$ (green) and $y=0$ (red) allow us to get a rectangular lattice.}
	\label{fig:hexgrid_hexgpu}
\end{figure}

For the medulla, we modeled the modular types identified in a dense 7 column (Figure \ref{fig:hexgrid_hexgpu} (left)) reconstruction by Takemura \etal \cite{Takemura2015}. These include Mi1, Mi4, Mi9, T2, T2a, T3, Tm20, Tm1, Tm2, Tm4, Tm9, TmY5a, Dm8, Dm2 which can be mapped once per column \cite{Takemura2015}. Additionally, we simulate T4a-T4d and T5a-T5d, which are ultraperiodic and connect with 4 distinct layers in the lobula plate \cite{Takemura2013}. A recent update by Takemura \etal \cite{Takemura2017} looked at T4 connections beyond 7 columns and found a better connectome model (Figure \ref{fig:fmap_new_T4a} versus \ref{fig:fmap_old_T4a} previously), which also significantly improved our simulation results (Section \ref{sec:results}). For T5 however, we know that Tm1, Tm2 and Tm9 synapse onto T5, but the spatial configuration and delays are not yet known, since T5 resides in the lobula. This also had consequences for recovering T5 functionality (Section \ref{sec:phys_resp}). Tm6 and Mi15 appear to be tiled since they were found only 4 times in 7 columns \cite{Takemura2013} (Figure \ref{fig:fullnet}), and we also simplified the ultraperiodic occurence of Tm3 to synperiodic. Many cells are transmedullar and project to the lobula or lobula plate \cite{Fischbach1989}. Transmedullar cells are the outputs of our simulated medulla and connect to a 3-stage fully connected decoder (Figure \ref{fig:fullnet}), with 128, 32 and 4 neurons, which integrates visual information to solve our proxy-task (Section \ref{sec:davis_tracking}).

Additionally, we used unpublished RNA sequence data to determine which neurons and synapses are inhibitory (such as Mi9) and excitatory, based on the neurotransmitters and receptors expressed. These constraints were also imposed during fine-tuning of the network weights (Section \ref{sec:backpropagation}). 

The electron microscopic reconstructions contain detailed information about the morphologies of the neurons being modeled. However, we ignore these details, instead choosing to build a simplified network of linear-nonlinear point neurons with continuous non-negative activations and instantaneous synapses. Since neurons at this stage of the fly visual system are non-spiking and produce graded potentials, this is a reasonable approximation.

We simulate a hexagonal lattice of $y=20$, $x=35$ point neurons with continuous activations, resulting in the typical $\sim 700$ ommatidia and retinotopic columns found in a \textit{Drosophila} fruit fly. Every neuron is defined as node with intrinsic properties: Name, sparsity along the $y$ and $x$-axis, activation function (ReLU), bias value and operator (addition). We selected the network orientation and reference frame (Figure \ref{fig:hexgrid_hexgpu} (left)) to be aligned with the dorsal half of the right \textit{Drosophila} eye and built the model so that we can disregard the optic chiasma between lamina and medulla.

\subsection{Connectome-based network weights}
\label{sec:network_weights}
Connections between neurons in our model are defined as edge between source and target neuron. All weights are replicated spatially for all neurons of the same type, like in regular convolution filters. Depending on how two neuron types are connected laterally, arbitrary sparse filters within hexagonal neighborhoods can arise (Figure \ref{fig:fmap_new_T4a}). 
In our model, we replaced all synapses counted between two connected distinct neurons in the connectome by a single synapse. The resulting synapse was then initialized with a weight equal to the number of counted synapses divided by an arbitrary normalization factor of $75$. This factor was chosen as small as possible, but large enough so that gradient-based optimization (Section \ref{sec:backpropagation}) was stable. Since we used leaky rectified linear units and a graded potential model (non-spiking) for all neurons, the choice of normalization factor is arbitrary and can be rescaled exactly by the activities of the neurons.
For the neuron bias values, there is no information that can be gained from the connectome and we therefore initialized them to a small positive constant value of $3.5/75.0$.

The complete model with all intercolumnar connection pattern and sparse filters can be found in the \textit{Supplemental Material}. In Figure \ref{fig:fullnet} we present our complete network graph. Each edge starts at the receptive field, which is marked in the same color as the edge, and projects to its target neuron. Note that the hexagonal lattice is shown as $15\times20$, whereas we used a $20\times35$ lattice for our simulation.

\subsection{Physiology-based synaptic delays}
\label{sec:synaptic_delays}
As we simulate our model as a convolutional neural network with recurrent components, the activity of every downstream neuron can only be computed when all inputs to a neuron are determined for step $t$. This means we needed to linearize the connections so that the network graph becomes a directed acyclic graph through time. For symmetric connections between neurons, autapses to the same neuron and feedback connections going back from the medulla to the lamina and ommatidia, we added a synaptic delay of $\SI{1}{{a.u.}}$, meaning the neurons read their input from state $t-1$, which is well-defined for the whole network from the previous forward pass.

\begin{figure}
	\centering
	\subfloat[T4a input filters according to the Takemura 2017 model \cite{Takemura2017}.]{
		\includegraphics[scale=0.69]{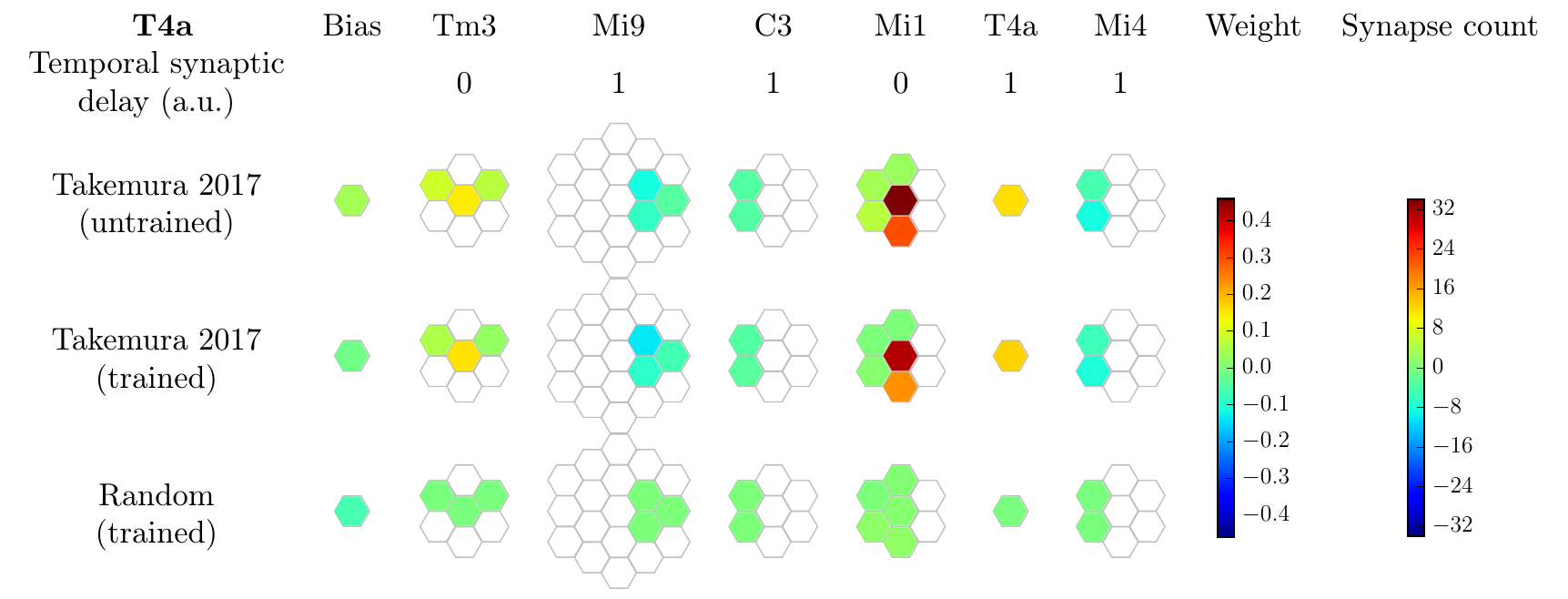}
		\label{fig:fmap_new_T4a}
	}
	\hspace{0mm}
	\subfloat[T4a input filters according to the Takemura 2013 model \cite{Takemura2013}.]{
		\includegraphics[scale=0.69]{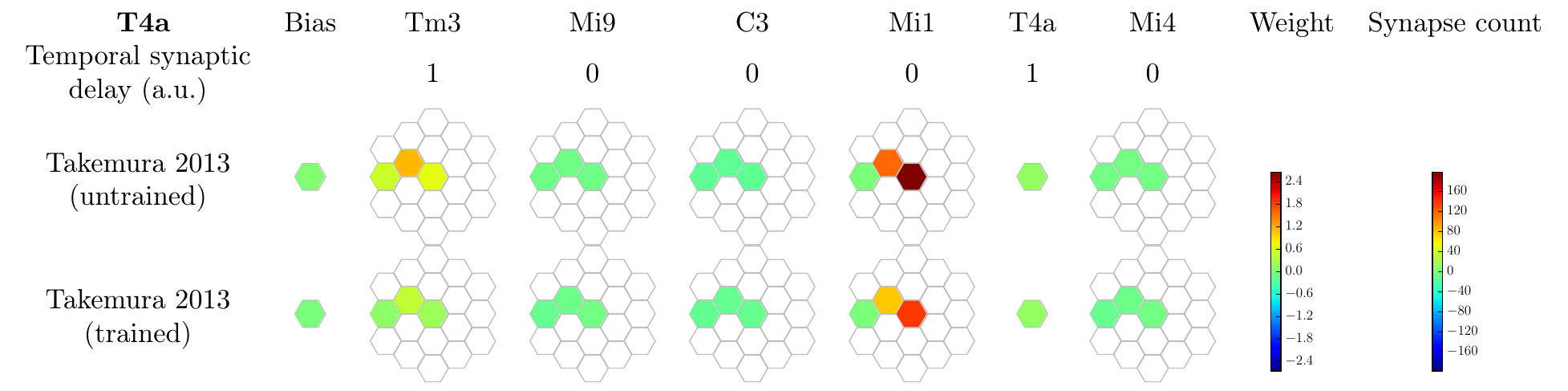}
		\label{fig:fmap_old_T4a}
	}
	\caption{Initial and emerging convolutional receptive field filters of T4a. Mi1 and Tm3 are the main excitatory contributions, while Mi9, C3 and Mi4 are inhibitory. Weights initialized from the Takemura 2017 connectome are similar to the initialization, however weights initialized randomly do not recover the connectome structure.}
\end{figure}

Additional synaptic delays were introduced where physiological data suggested necessary delays: For T4, because of the temporal frequency optimum \cite{Maisak2013} and the spatial offsets of Mi1 to Tm3 \cite{Takemura2015} and only small temporal delay between them \cite{Behnia2014}, we expect a delay at the negative input \cite{Takemura2017} (Figure \ref{fig:fmap_new_T4a}), which would form a more complex type of BL \cite{Barlow1965} detector with additional properties not described by such simple models (Section \ref{sec:phys_resp}). Our previous model based on older data \cite{Takemura2013} suggested a configuration more akin to a HR \cite{Reichardt1961} model. For the T5 cells, due to the lack of a complete connectome, we assumed the same configuration of temporal and spatial offsets.

\subsection{Implementation}
In order to simulate our model efficiently, we used the tensor data structures and gradient based optimizers from the Caffe deep learning library \cite{Caffe2017} combined with runtime code generating modules from LibDNN \cite{LibDNN2017}. We introduce a new layer, \textit{sparse repeated pattern recurrent neural network}, which takes care of translating our connectome-based network defined in Python to runtime generated CUDA or OpenCL code. Caffe's Python interface allows to conduct experiments rapidly, as it can be combined with Python tools to generate artificial stimuli and analyze the neuronal responses.

The newly introduced layer supports backpropagation through time (BPTT, \cite{Mozer1995}\cite{Robinson1987}) during training, and a variable history length can be chosen. In our setup, we used a batchsize of 5 steps and a history of 10 steps. Images have to be resampled and rearranged for the hexagonal lattice (Figure \ref{fig:hexgrid_hexgpu} (right)). To this end, we implemented nearest neighbor and bilinear interpolations as well as an ommatidia-mode which resamples an RGB image to grayscale for the rhabdomeres R1-R6, which all receive the same input in our model, R7 receives red (simulates UV), $30\%$ of R8 sample blue and $70\%$ of R8 sample green \cite{Morante2004}. Note that we normalize the images and do not simulate any spherical projections that occur in real fly vision \cite{Lazar2015}, since our focus is on the lamina and medulla circuitry.

\begin{figure}
	\includegraphics[width=0.95\linewidth]{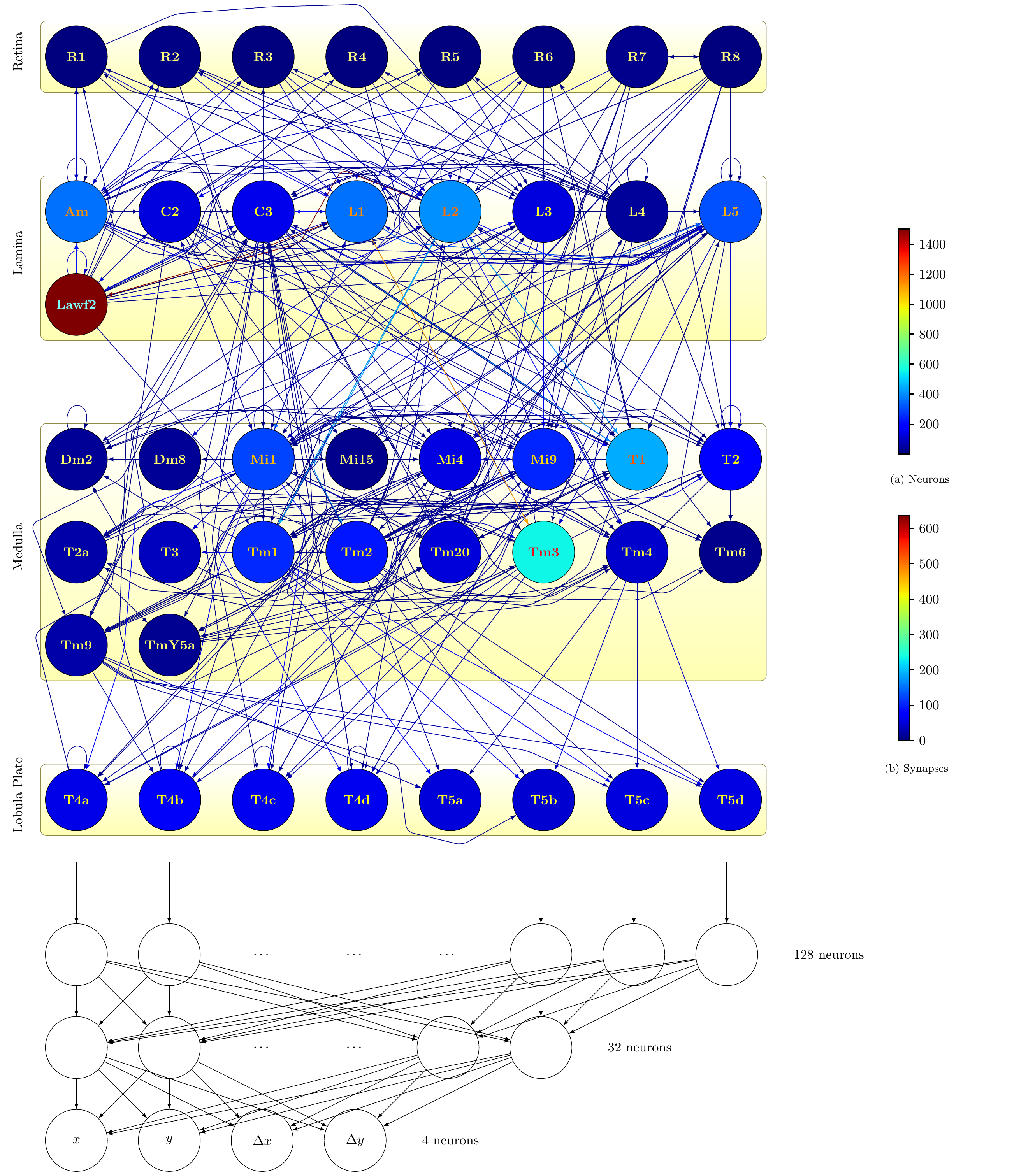}
	\caption{\textit{Drosophila} visual system model based on \cite{Takemura2017} with a 3 layer decoder, displaying the total number of input synapses to a neuron (a) and the synapses per neuron type-to-type edge (b).}
	\label{fig:fullnet_ut}
\end{figure}

\begin{figure}
	\includegraphics[width=0.95\linewidth]{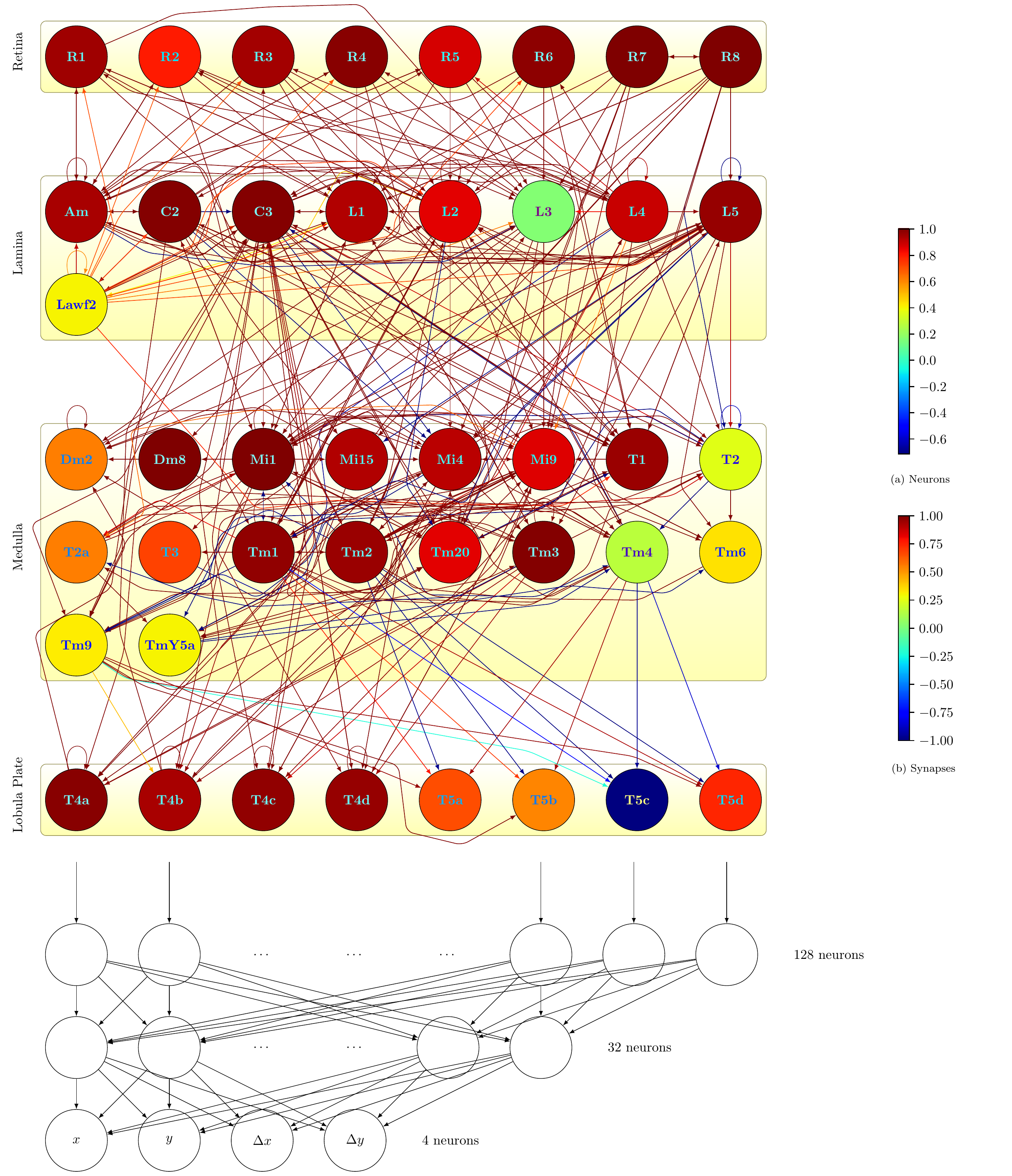}
	\caption{Correlation coefficient computed between trained weights and the connectome initialized weights based on \cite{Takemura2017} show that the on pathway weights are largely unchanged. Significant weight changes occur largely in the off pathway for which high quality connectome data was not available.}
	\label{fig:fullnet_t_graph}
\end{figure}

\begin{figure}
	\includegraphics[width=0.95\linewidth]{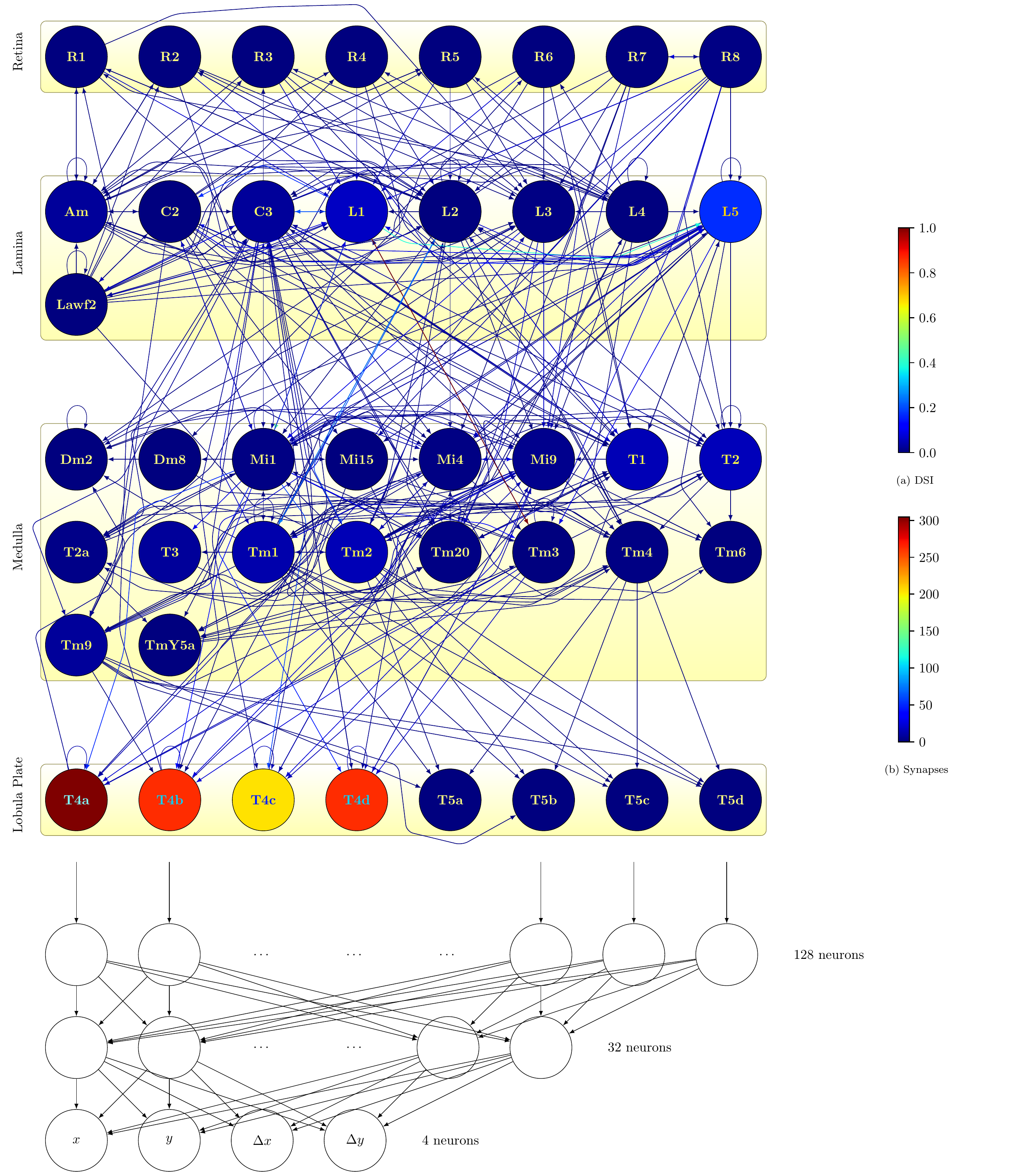}
	\caption{Takemura 2017 initialized and trained network shows strong direction selectivity arising in the T4 neurons but not in their inputs, consistent with known physiology. Cell type nodes are color-coded by the direction selectivity index (DSI) of neurons (a) and the synapses per neuron type-to-type edge (b).}
	\label{fig:fullnet_t_dsi_graph}
\end{figure}

\section{Gradient-based circuit optimization for object tracking}
\label{sec:optimization}
In order to fine-tune the weights of our networks, we used a proxy-task that depends on the circuit's ability to compute a moving object's position and velocity. This approach is in contrast to explicitly training the network to learn an encoding model based on physiologically measured neural responses \cite{Lee2017}, and thus requires only anatomical information, and no recordings of neural activity.

\subsection{DAVIS dataset}
\label{sec:davis_dataset}
We used the DAVIS dataset \cite{DAVIS2016,DAVIS2017} in its 2016 version with 480p image resolution, which is used as a video object segmentation challenge. Due to its dense single-object annotation, it is easy to compute ground truth for the object centroid $(x,y)$ and velocity $(\Delta x,\Delta y)$ between consecutive frames. The network's task is to predict these four parameters.
The whole dataset consists of 50 independent sequences of 40 to 100 frames at $\SI{24}{FPS}$. For our purposes, we selected 8 sequences (\textit{bmx-bumps}, \textit{drift-chicane}, \textit{motorbike}, \textit{paragliding}, \textit{rollerblade}, \textit{soccerball}, \textit{surf} and \textit{kite-walk}) that contain primarily object motion and maintain a static camera position. Other sequences were not suitable, as the camera follows the object too closely, leading to near-zero object motion relative to the camera.
The selected sequences were split into blocks of 10 consecutive frames. Per set, 2 blocks of 10 frames were used for testing, and the remaining 2 to 8 blocks were used as training set. Additionally, we augmented the dataset by mirroring the training data along the vertical and horizontal axis.

\subsection{Gradient based optimization of network parameters}
\label{sec:backpropagation}

During circuit optimization, we used the Adam optimizer \cite{Adam2015} with weight decay to avoid large weights. The 164 decoder neurons were always initialized with Xavier \cite{Glorot2010} weights without constraints. For our connectome-based network, six different training configurations were tested for functionality and robustness.

\begin{itemize}[noitemsep]
	\item {\bf Takemura 2017 (trained)} Sign constrained weights initialized from the connectome.
	\item {\bf Takemura 2017 $\cdot$ 40\% noise (trained)} Initial weights perturbed with $40\%$ uniformly sampled multiplicative noise.
	\item {\bf Takemura 2017 (untrained)} Weights fixed to connectome, with only biases and decoder weights trained.
	\item {\bf Random (trained)} Sign constrained weights initialized at random with MSRA \cite{He2015}.
	\item {\bf Random $\cdot$ 40\% noise (trained)} Sign constrained weights using the previous MSRA initialization with $40\%$ uniformly sampled multiplicative noise.
	\item {\bf Takemura 2013 (trained)} Sign constrained weights using our older model \cite{Takemura2013} for T4 inputs (Figure \ref{fig:fmap_new_T4a}).
\end{itemize}

For the random (MSRA \cite{He2015}) initialized weights, the sign constraints were taken from the connectome initialized weights, since we only want to consider models which keep known inhibitory and excitatory properties. Bias values are not limited since their initial values are arbitrary (Section \ref{sec:network_weights}). The weight constraints were enforced with a projected gradient descent update, $\theta_t' = \sign(\theta_0) \cdot |\theta_t|$, that mirrors the weights on the $0$-axis rather than setting them to $0$, where $\theta_t$ is the weight obtained by Adam \cite{Adam2015} optimization in step $t$. This keeps as many weights from being disabled as possible.
Each configuration was trained for 30,000 iterations to minimize the squared error in predicting object location and velocity.

\section{Results}
\label{sec:results}

\subsection{DAVIS object tracking}
\label{sec:davis_tracking}
\begin{figure}
	\centering
	\subfloat{
	\includegraphics[scale=0.5]{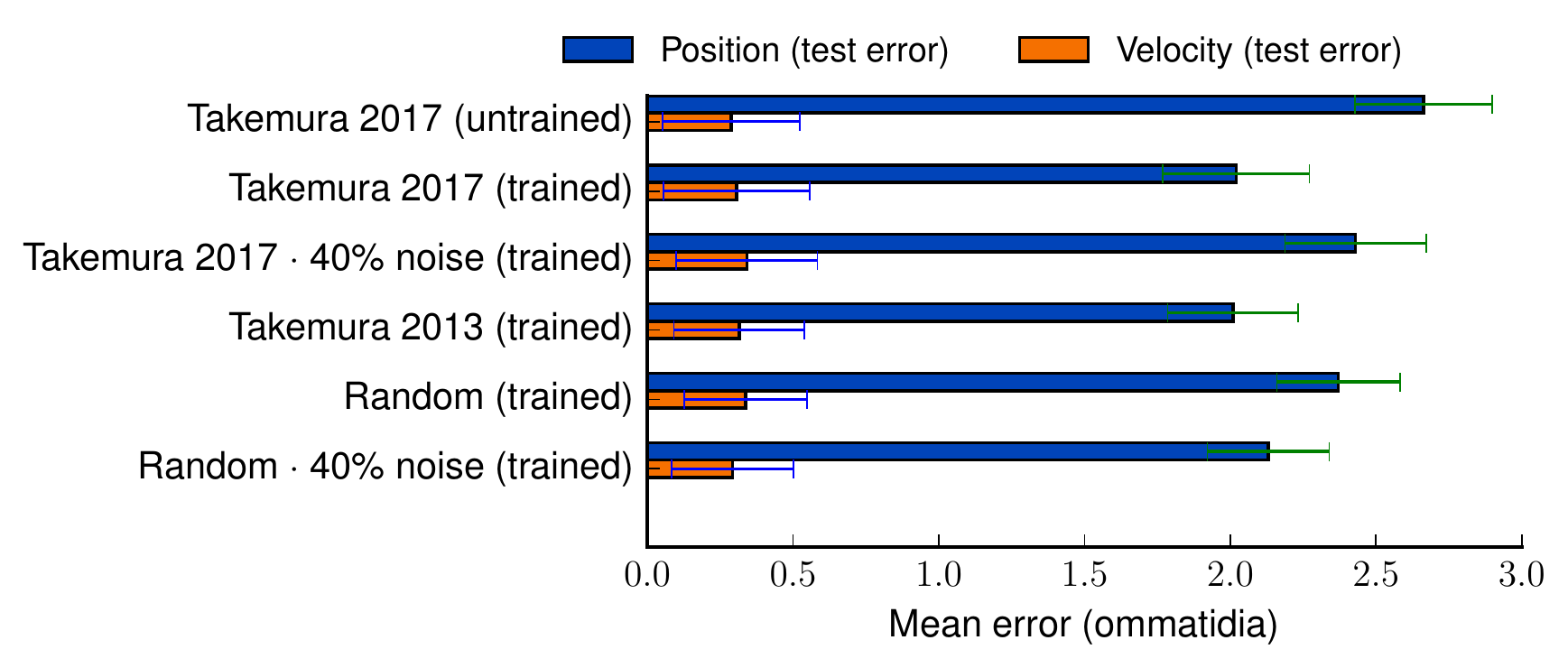}
	}
% Davis test results:
% [ 0.29240702  0.33728912  0.31545151  0.34144386  0.30616091  0.28833019]
% [ 2.13043717  2.37123344  2.00911625  2.42982701  2.02030096  2.66305113]
	\hspace{10mm}
	\centering
	\subfloat{
	\includegraphics[scale=0.5]{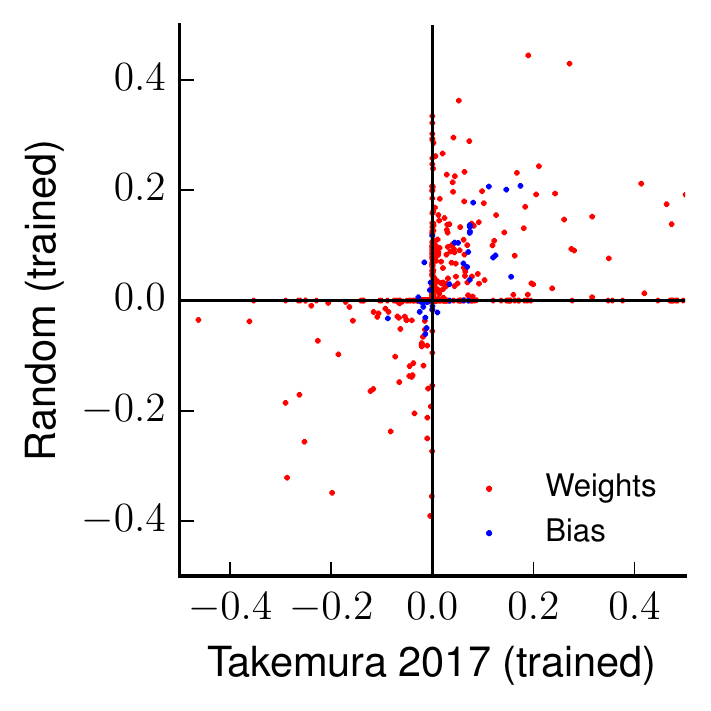}
	}
	\caption{(left) Test scores (mean $L_2$ error) on the DAVIS dataset based on 20 frames for each of the selected 8 sequences in the DAVIS dataset (Section \ref{sec:davis_dataset}). (right) Different weight and bias solutions are discovered by training from connectome vs random initialization.}
	\label{fig:davis_error_scatter}
\end{figure}

Our networks are able to track objects with a mean $L_2$ error as low as $\SI{2.0}{\text{ommatidia}}$ in  distance on the test set. The network which was constrained to only train the decoder and the bias values of the network was found to perform worst, with $\SI{2.6}{\text{ommatidia}}$ error (Figure \ref{fig:davis_error_scatter} (left)). Predicted velocities had a mean $L_2$ error within a range of $0.28$ to $\SI{0.33}{\text{ommatidia/step}}$.

\subsection{Predicted neural tuning properties}
\label{sec:phys_resp}
\begin{figure}
	\centering
	\includegraphics[width=0.95\linewidth]{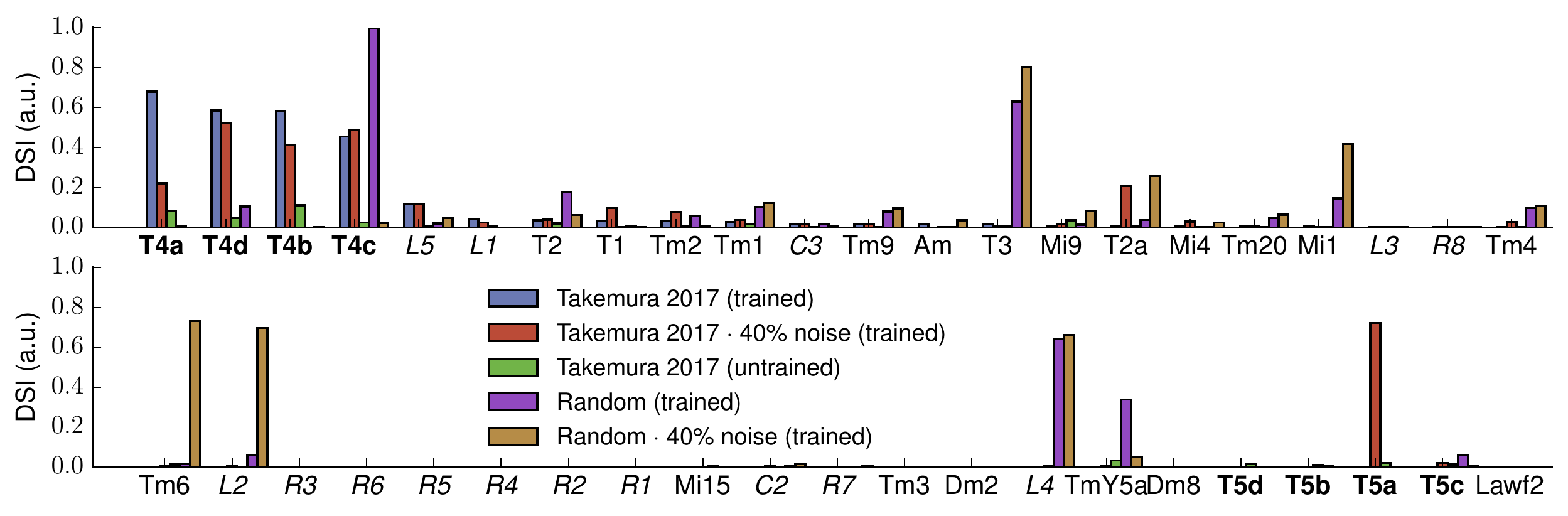}
	\caption{Direction selectivity index (DSI) scores. \textbf{Bold} neurons are known to be direction selective, while \textit{italic} neurons are known to be non-selective. The remaining neurons have unknown tuning properties. The Takemura 2017 model successfully recovers \textbf{T4a-T4d} direction selectivity, while random initialization does not.}
	\label{fig:dsi}
\end{figure}

We recorded all cell types both qualitatively as video clip (see \textit{Supplemental Material}) and quantitatively by checking for linear and nonlinear behavior. Most notably, we found that direction and orientation selective properties were recovered during training (Section \ref{sec:backpropagation}).
To compute the direction selectivity index (DSI) and orientation selectivity of all 43 neuron types in our simulation (see Figure \ref{fig:dsi}), we moved light and dark bars of 180 directions across the visual field of our simulated fly eye and measured the peak responses $R_{k}^{peak}$. For direction selectivity, the bars had a width of $\sim\SI{1}{\text{ommatidia}}\,(\estimates 5^\circ)$ and a speed of $\sim\SI{0.7}{\text{ommatidia/timestep}}$. Orientation selectivity was determined with a bar of $\SI{2}{\text{ommatidia}}$ width and a speed of $\sim\SI{0.1}{\text{ommatidia/step}}$, which is slightly below the threshold for direction selectivity.

A robust method based on vector addition \cite{Mazurek2014} was used to determine a noise-free selectivity index: $DSI = \sqrt{(\sum_{k=0}^{2\pi}\cos(k)\cdot R_{k}^{peak})^2+(\sum_{k=0}^{2\pi}\sin(k)\cdot R_{k}^{peak})^2}$

\begin{figure}
	\centering
	\begingroup
	\captionsetup[subfigure]{width=55mm}
	\subfloat[T4a NULL response ($0^{\circ}$).]{
	\includegraphics[width=53mm]{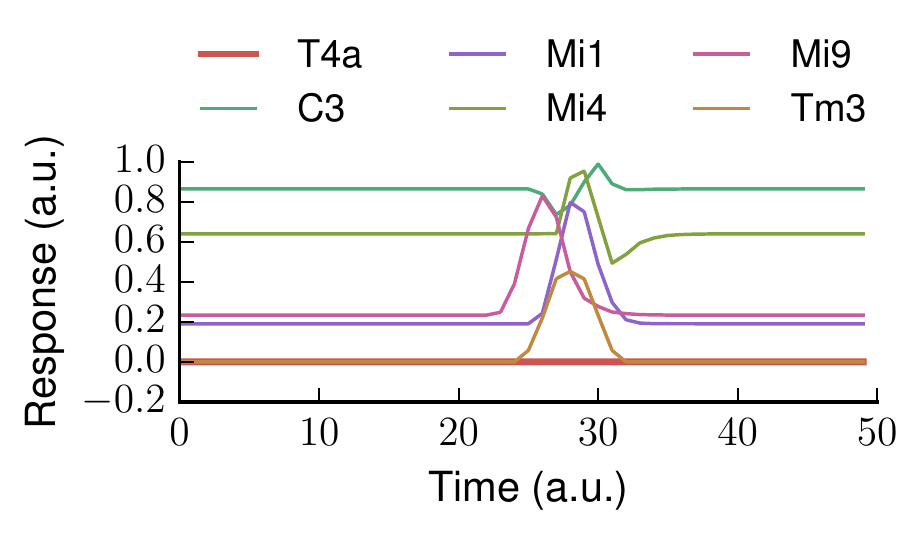}
	\label{fig:T4a_null}
	}
	\hspace{10mm}
	\subfloat[T4a preferred direction ($192^{\circ}$) response.]{
	\includegraphics[width=53mm]{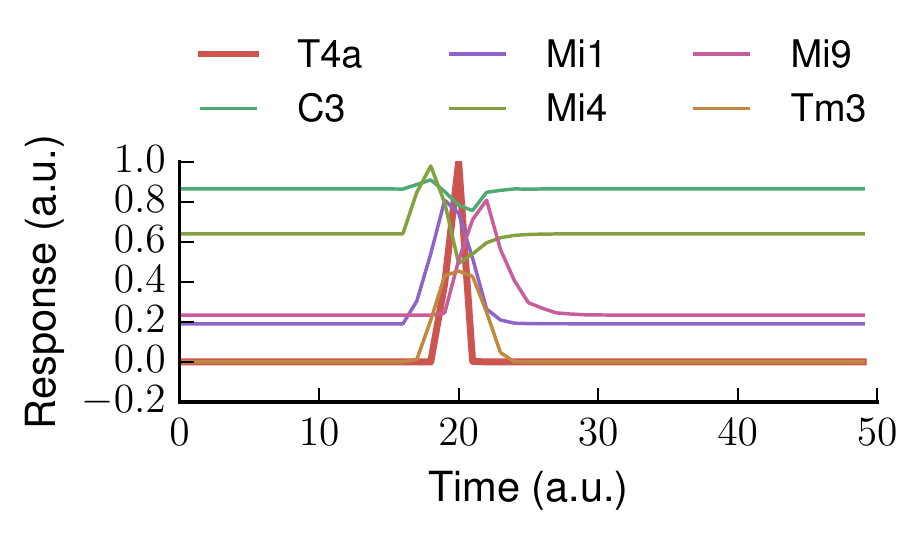}
	\label{fig:T4a_preferred}
	}
	\hspace{10mm}
	\subfloat[DS, Takemura 2017 (untrained).]{
	\includegraphics[width=67mm]{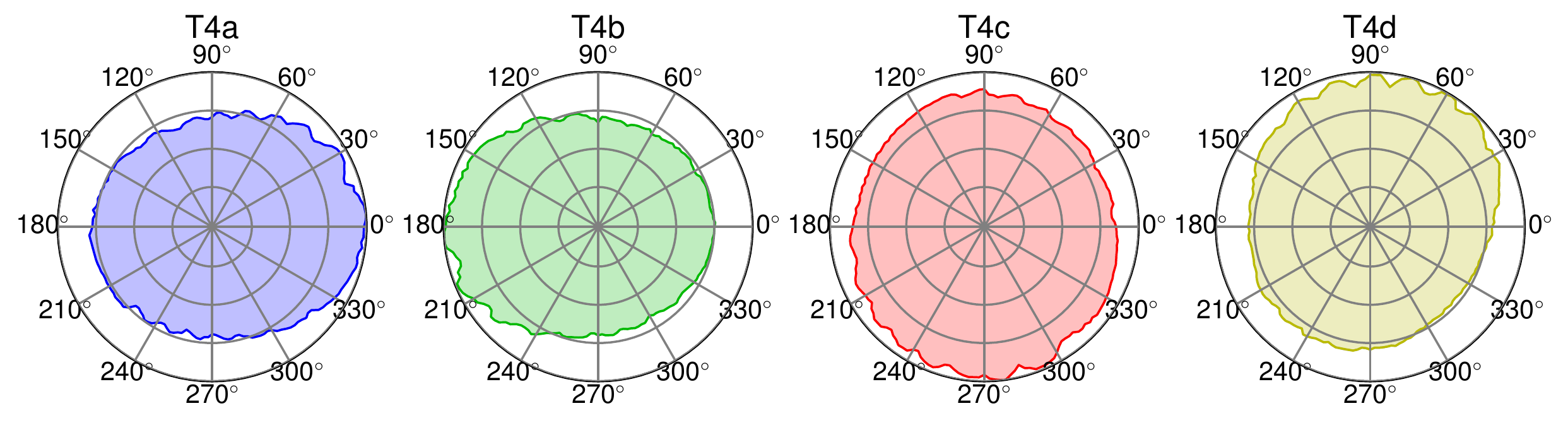}
	\label{fig:DS2017ut}
	}
	\hspace{0mm}
	\subfloat[OS, Takemura 2017 (untrained).]{
	\includegraphics[width=67mm]{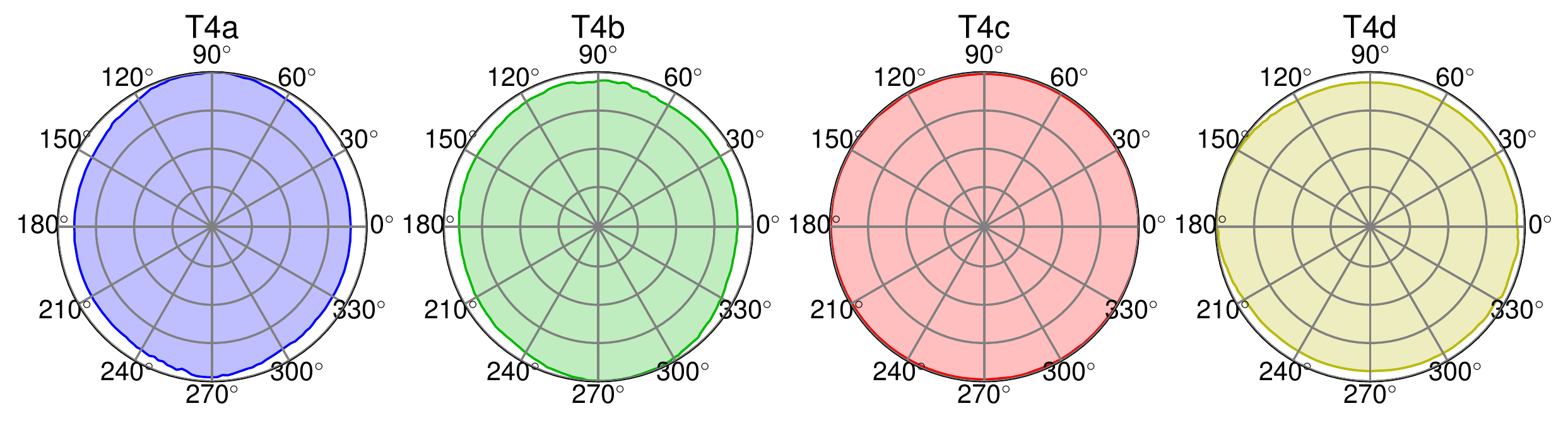}
	\label{fig:OS2017ut}
	}
	\hspace{0mm}
	\subfloat[DS, Takemura 2017 (trained). Each T4 cell is direction selective in one of four principal directions (T4a: $192^\circ$, T4b: $359^\circ$, T4c: $51^\circ$ and T4d: $275^\circ$).]{
	\includegraphics[width=67mm]{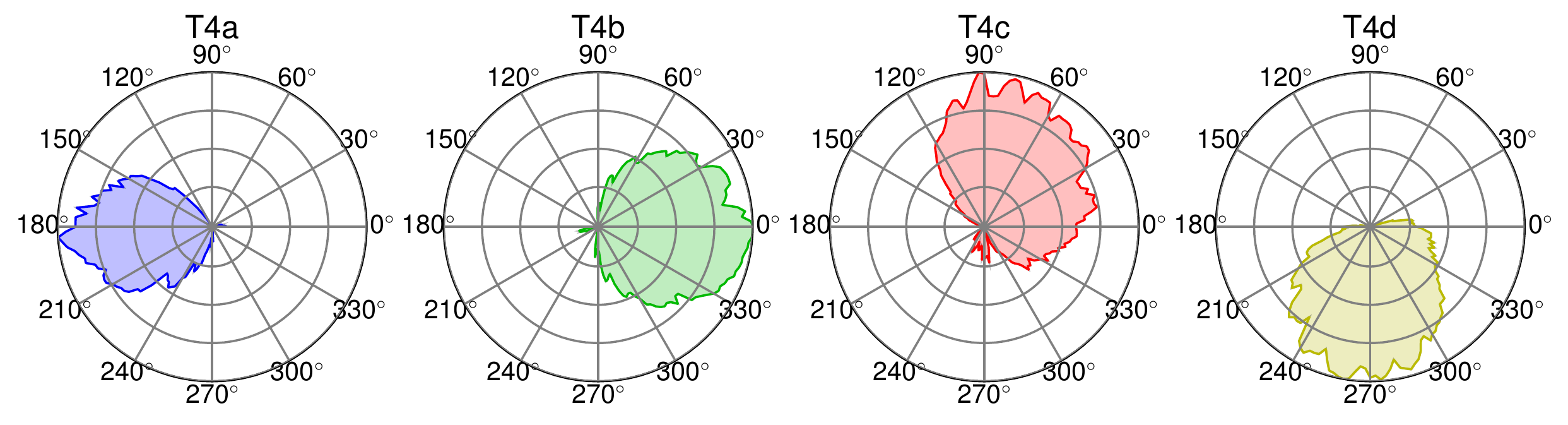}
	\label{fig:DS2017t}
	}
	\hspace{0mm}
	\subfloat[OS, Takemura 2017 (trained). Two-lobed orientation selectivity with strong center-surround inhibition for bars orthogonal to the preferred direction as shown by \cite{Fisher2015}.]{
	\includegraphics[width=67mm]{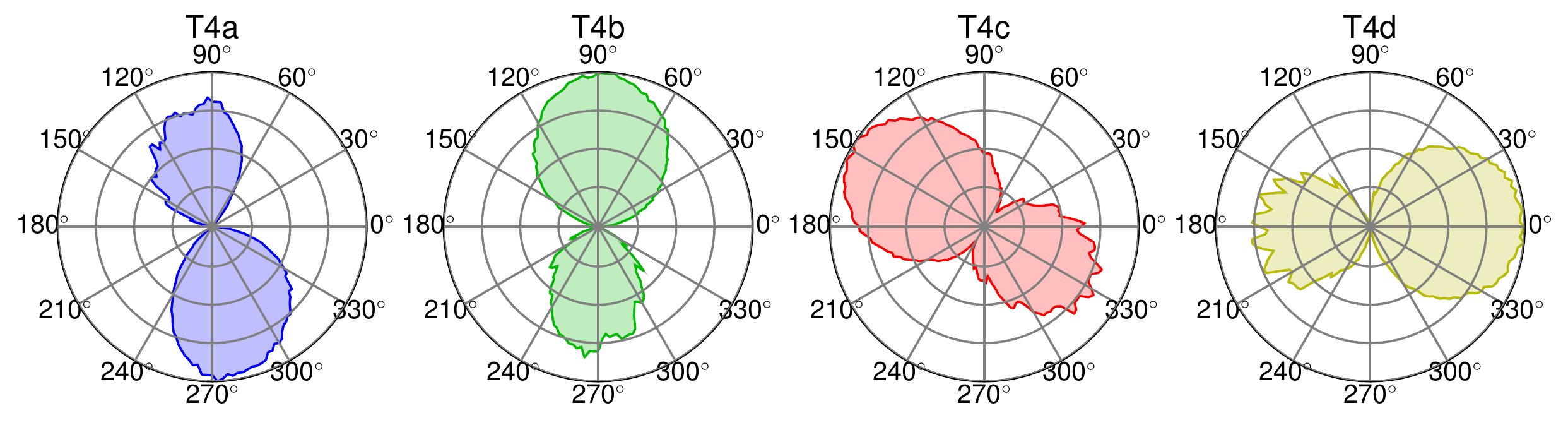}
	\label{fig:OS2017t}
	}
	\hspace{0mm}
	\subfloat[DS, Random (trained).]{
	\includegraphics[width=67mm]{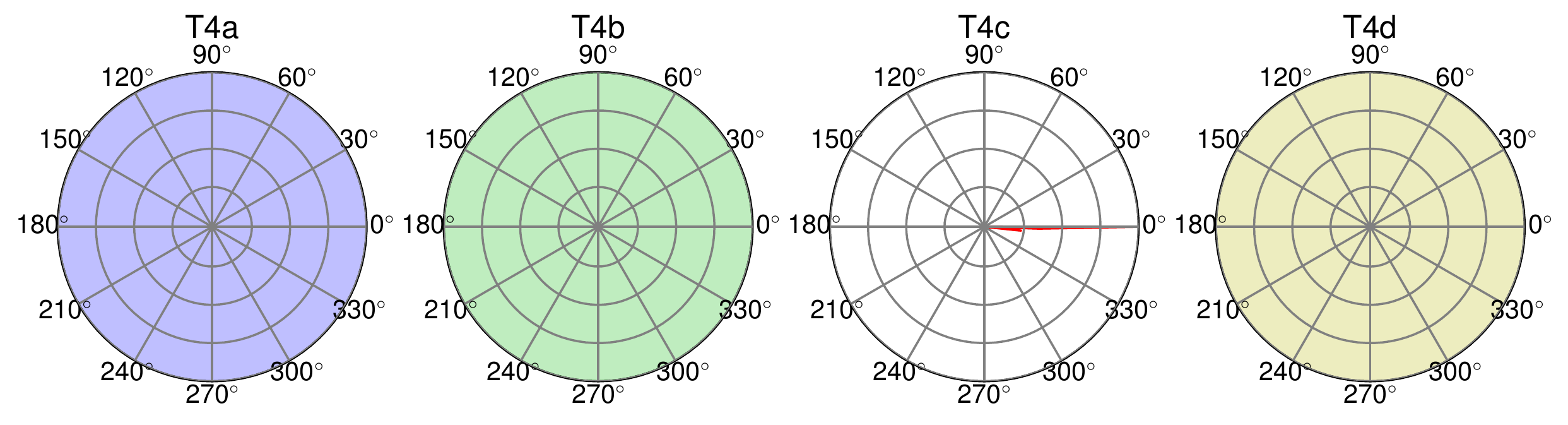}
	\label{fig:DSrandomt}
	}
	\hspace{0mm}
	\subfloat[OS, Random (trained).]{
	\includegraphics[width=67mm]{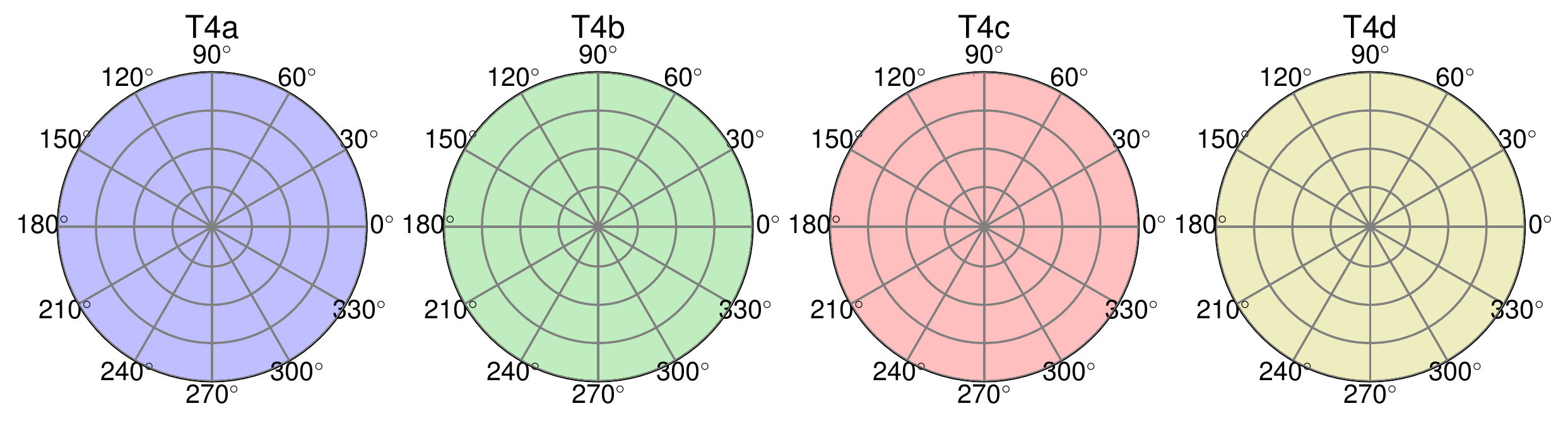}
	\label{fig:OSrandomt}
	}
	\hspace{0mm}
	\subfloat[DS, Takemura 2013 (trained).]{
	\includegraphics[width=67mm]{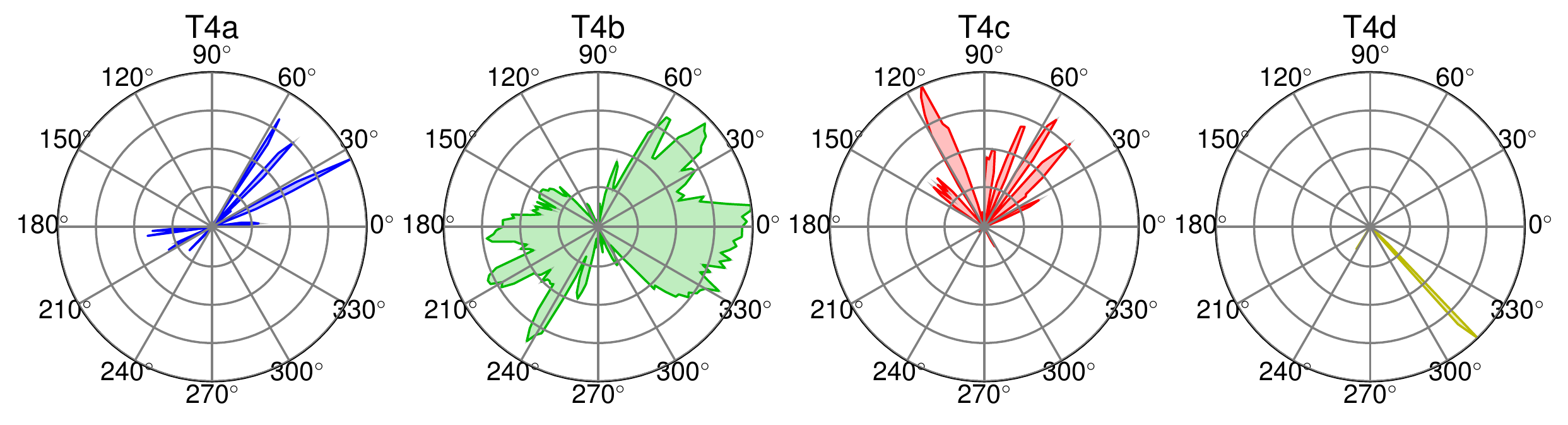}
	\label{fig:DS2013t}
	}
	\hspace{0mm}
	\subfloat[OS, Takemura 2013 (trained).]{
	\includegraphics[width=67mm]{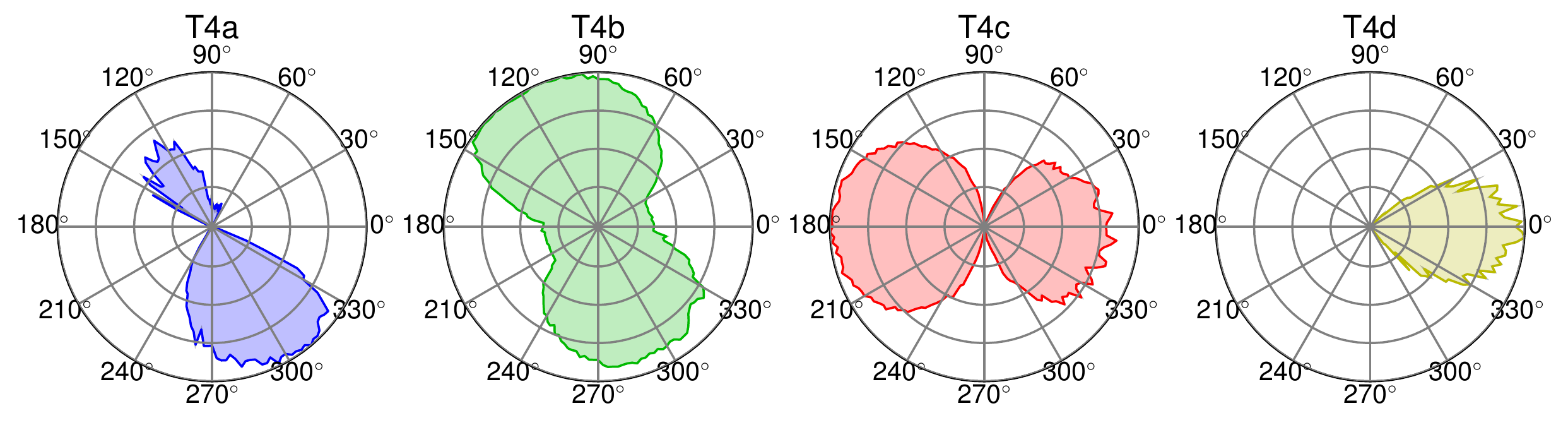}
	\label{fig:OS2013t}
	}
	\endgroup
	\caption{Orientation (OS) and direction (DS) selectivity of T4 cells. Responses were tested with light and dark bars of 180 directions.}
\end{figure}

We found that training the Takemura 2017 initialized weight model with polarity constraints obtained from connectome data recovers and tunes T4 cells to well-known physiological responses \cite{Fisher2015,Maisak2013,Serbe2016}. The neurons became highly direction selective in their preferred orientation (T4a: $192^\circ$, T4b: $359^\circ$, T4c: $51^\circ$ and T4d: $275^\circ$, Figure \ref{fig:DS2017t}, angles are oriented as in Fisher's measurements \cite{Fisher2015}). They also show a two-lobed orientation selectivity with strong center-surround inhibition for bars orthogonal to the preferred direction (Figure \ref{fig:OS2017t}). Previous models \cite{Takemura2013} were also able to tune the cells, although less stable (Figures \ref{fig:DS2013t}-\ref{fig:OS2013t}). Networks initialized with random weights were unable to recover these properties (Figures \ref{fig:DSrandomt}-\ref{fig:OSrandomt}). The characteristic tuning properties were also absent with initial weights, however slight DS tuning is visible (Figures \ref{fig:DS2017ut}-\ref{fig:OS2017ut}).

\subsection{Robustness of connectome-initialized networks}
Although networks initialized with random weights were also able to track objects (Section \ref{sec:davis_tracking}), the circuitry recovered to achieve this does not correspond to physiology, unlike networks initialized with synapse counts. The high disparity between connectome-based and random weights after training (Figure \ref{fig:davis_error_scatter} (right)) are also an indicator that the network has many local minima to solve the task, as predicted by the universal function approximation \cite{Csaji2001,Cybenko1992}. We found that random weights trained networks also recovered DS responses for some neurons (Figure \ref{fig:dsi}) at sufficiently dissimilar angles (T3: $10^\circ$ and L4: $238^\circ$).
Solutions initialized at random are also very unstable, and adding $40\%$ multiplicative noise to the initialization leads to different neurons becoming direction selective (such as L2 and Tm6 instead of T4c and TmY5a). This also shows that many neurons have the spatiotemporal wiring required to potentially become direction selective.

Our Takemura 2017 \cite{Takemura2017} model with connectome-based weights, on the other hand, is robust against up to $40\%$ multiplicative noise, at which point it still recovers identical, although less sharply tuned, responses for T4a-T4d. Additionally, it enabled T5a to become direction selective, although at the wrong angle ($335^\circ$ instead of the expected $\sim 180^\circ$).

\section{Discussion}
\label{sec:discussion}
Our model predicted accurate responses for all T4 cells and demonstrates that functional information can be gained by means of fine-tuning a model initialized with weights derived from electron microscopic reconstructions with only minimal additional assumptions about synaptic delays. Remarkably, the model converged to such responses without being forced to learn them, as in encoding models \cite{Lee2017}, but instead through a visual proxy-task with 4 predicted variables.

We acknowledge that not all neurons took on their expected ON/OFF responses suggested in the literature. While for example Mi1, Tm3 and Mi4 (ON) and Tm1, Tm2, Tm4, Tm9 (OFF) are responding as expected, Mi9 shows ON characteristics instead of OFF (as visualized in the video clip in our \textit{Supplemental Material}). Many parts of our model still contain uncertainties and incomplete filters due to missing connectomic information and the difficulty of assembling a consensus-model across multiple publications. Examples include the T5 pathway, for which we were unable robustly recover OFF direction selectivity, likely due to incomplete lateral dendritic arbor reconstructions. Indeed, the T4 neurons only recovered their known direction and orientation selectivity when their connectome was reconstructed with sufficient accuracy, as in \cite{Takemura2017}. The incomplete reconstructions from \cite{Takemura2013} were not sufficient. Note also that neurons with large tangential branches such as Mt8, Mi11 or Tm28 within the medulla were also not included in our model \cite{Fischbach1989}.

We also anticipate that our model could be useful to also test later stages of fly vision, such as vertical and horizontal system cells \cite{Borst2010}, object-detecting neurons such as LC11 \cite{Keles2017} and visual projection neurons \cite{Wu2016,Otsuna2006}. These steps await further dense EM reconstructions of the lobula and lobula plate. With more complete models, we expect to be able to extract further characteristics such as object detection, color vision and rotational movement sensitivity from structure alone, which could then be verified with physiological measurements or be used to make predictions. Furthermore, we hope to automate the reconstruction of models that can be simulated directly from EM reconstructions without the need to assemble coherent networks from various publications.

% Commented for anonymous submission
%\subsubsection*{Acknowledgments}
%
%The authors would like to thank Dr. Michael Reiser for helpful insights into inhibitory and excitatory pathways within the lobula and medulla of the \textit{Drosophila's} visual system.

% Put refs on its own page (9th page) for NIPS
\newpage
\section*{References}
\small

%\nocite{*}
\printbibliography[heading=none]

\end{document}